\documentclass[sigconf]{acmart}
\usepackage{multirow}
\usepackage{url}
\usepackage{cleveref}
\usepackage{xcolor}
\usepackage{tcolorbox}
\usepackage{subcaption}
\usepackage{color}
\usepackage{bbding}

\AtBeginDocument{%
  }

\setcopyright{acmlicensed}
\copyrightyear{2025}
\acmYear{2025}
\begin{document}

% \title{LLM4MEA: Data-free Model Extraction Attacks on\\ Sequential Recommenders via Large Language Models}
% for arxiv:
\title{LLM4MEA: Data-free Model Extraction Attacks on\\ Sequential Recommenders via Large Language Models}

\author{Shilong Zhao}
\email{zhaoshilong23s@ict.ac.cn}
\orcid{0009-0001-8420-0311}
\affiliation{
  \institution{State Key Lab of AI Safety, Institute of Computing Technology, CAS\\ University of Chinese Academy of Science}
  \city{Beijing}
  \country{China}
}

\author{Fei Sun}
\authornote{Corresponding author.}
\email{sunfei@ict.ac.cn}
\affiliation{
  \institution{State Key Lab of AI Safety, Institute of Computing Technology, CAS}
  \city{Beijing}
  \country{China}
}

\author{Kaike Zhang}
\email{zhangkaike21s@ict.ac.cn}
\affiliation{
  \institution{State Key Lab of AI Safety, Institute of Computing Technology, CAS\\ University of Chinese Academy of Science}
  \city{Beijing}
  \country{China}
}

\author{Shaoling Jing}
\email{jingshaoling@ict.ac.cn}
\affiliation{
  \institution{State Key Lab of AI Safety, Institute of Computing Technology, CAS}
  \city{Beijing}
  \country{China}
}

\author{Du Su}
\email{sudu@ict.ac.cn}
\affiliation{
  \institution{State Key Lab of AI Safety, Institute of Computing Technology, CAS}
  \city{Beijing}
  \country{China}
}

\author{Zhichao Shi}
\email{shizhichao22s@ict.ac.cn}
\affiliation{
  \institution{State Key Lab of AI Safety, Institute of Computing Technology, CAS\\ University of Chinese Academy of Science}
  \city{Beijing}
  \country{China}
}

\author{Zhiyi Yin}
\authornotemark[1]
\email{yinzhiyi@ict.ac.cn}
\affiliation{
  \institution{State Key Lab of AI Safety, Institute of Computing Technology, CAS}
  \city{Beijing}
  \country{China}
}

\author{Huawei Shen}
\email{shenhuawei@ict.ac.cn}
\affiliation{
  \institution{State Key Lab of AI Safety, Institute of Computing Technology, CAS}
  \city{Beijing}
  \country{China}
}

\author{Xueqi Cheng}
\email{cxq@ict.ac.cn}
\affiliation{
  \institution{State Key Lab of AI Safety, Institute of Computing Technology, CAS}
  \city{Beijing}
  \country{China}
}
\renewcommand{\shortauthors}{Zhao, et al.}

\begin{abstract}
Recent studies have demonstrated the vulnerability of sequential recommender systems to Model Extraction Attacks (MEAs).
MEAs collect responses from recommender systems to replicate their functionality, enabling unauthorized deployments and posing critical privacy and security risks.
Data-free attacks in prior MEAs are ineffective at exposing recommender system vulnerabilities due to random sampling in data selection, which leads to misaligned synthetic and real-world distributions.
To overcome this limitation, we propose LLM4MEA, a novel model extraction method that leverages Large Language Models (LLMs) driven agent to generate data. 
It generates data through interactions between the agent and target recommender system.
In each interaction, the agent 
analyzes historical interactions to understand user behavior,
and selects items from recommendations with consistent preferences to extend the interaction history, which serves as training data for MEA.
Extensive experiments demonstrate that LLM4MEA significantly outperforms existing approaches in data quality and attack performance.
We also assess how the target system’s hyperparameters affect MEA and suggest a simple defense to reduce these risks.
The aim of this work is to raise awareness of the security and privacy risks of MEAs in recommender systems.
\end{abstract}

\begin{CCSXML}
<ccs2012>
   <concept>
       <concept_id>10002951.10003317.10003347.10003350</concept_id>
       <concept_desc>Information systems~Recommender systems</concept_desc>
       <concept_significance>500</concept_significance>
       </concept>
   <concept>
       <concept_id>10002978.10003022.10003028</concept_id>
       <concept_desc>Security and privacy~Domain-specific security and privacy architectures</concept_desc>
       <concept_significance>500</concept_significance>
       </concept>
 </ccs2012>
\end{CCSXML}

\ccsdesc[500]{Information systems~Recommender systems}
\ccsdesc[500]{Security and privacy~Domain-specific security and privacy architectures}

\keywords{Model Extraction Attack, Sequential Recommendation, Large Language Model, Agent, Data Synthesis}


\maketitle

\section{Introduction}
Sequential recommender systems are widely used in various domains like e-commerce and social media.
However, recent studies~\cite{yue2021black, zhang2024few, zhu2023model} have demonstrated the vulnerability of sequential recommender systems to Model Extraction Attacks (MEAs).
MEAs aim to train a surrogate model that mimics the behavior of the original recommender systems.
They user sequential data and its recommendations to train a surrogate model, which replicates the original system's behavior.
The extracted substitute models enable unauthorized model deployment~\cite{zhu2023model, zhang2024few, yue2021black} or downstream white-box attacks~\cite{zhu2023membership, yue2021black}.
It poses serious privacy threats and security risks to recommenders.

\begin{figure}[]
  \centering
  \includegraphics[width=\linewidth]{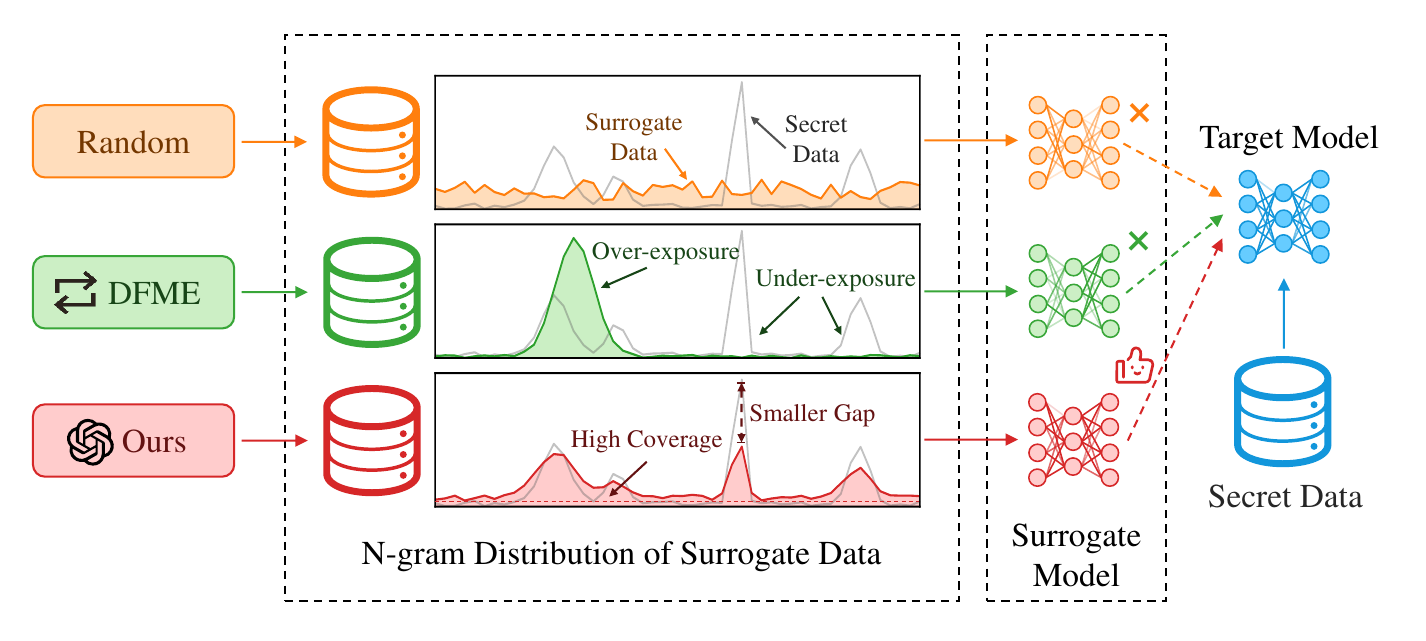}
  \caption{The MEA framework trains surrogate models on synthetic data via different methods (Random, SR-DFME~\cite{yue2021black} and Ours). Random data lacks specific patterns, SR-DFME suffers from over/under-exposure. Our method employs an LLM-driven Agent and debiasing to produce data with broader coverage and reduced gap, yielding superior MEA performance.}
  \label{fig:intro0}
\end{figure}

However, in real-world scenarios, accessing original training data is challenging because it is protected by strict authorization restrictions. 
With data-free setting, 
existing methods typically use public data or synthetic data for MEAs.
Yet using public data~\cite{zhao2023extracting, karmakar2023marich, jindal2024army, peng2023query} is unsuitable for MEA targeting sequential recommenders due to incompatible item vocabularies across different domains and platforms.
Recent works~\cite{shao2023data,yue2021black,wang2023data, liuimportance} address this issue through data synthesis, where user sequential histories are either randomly generated or constructed autoregressively with random sampling.
However, using random sampling as decision mechanism for data generation leads to neglecting both sequential patterns and preference consistency.
As shown in~\Cref{fig:intro0}, this results in significant distribution mismatches between synthetic and original training data: random generation lacks specific patterns, and SR-DFME's autoregressive sampling leads to over/under-exposure of items, ultimately limiting attack performance.

As Large Language Models (LLMs) have shown significant intelligence and remarkable capabilities in various scenarios including standalone tasks and human interaction~\cite{xi2023rise, li2023metaagents}, we leverage them to address the limitations of random sampling in MEAs.
With their ability to simulate human-like behavior, LLMs can potentially maintain consistent preferences while understanding and responding to recommendations naturally.
This insight inspires us to leverage LLMs as intelligent agents in the data generation process to generate more realistic data.

Based on above insights, we propose LLM4MEA, a novel model extraction method that leverages LLM driven agent to generate data. 
Specifically, our method generates training data through interactions between the LLM-driven agent and target recommender system. In each round, the agent comprehends historical interactions and selects items from recommendations based on consistent preferences, generating high-quality training data for MEAs.

However, when processing lengthy historical items, LLMs face not only significant computational overhead, but also the challenge of fully utilizing information within the long context
~\cite{an2024make, li2024long}.
To overcome these limitations, we introduce the Memory Compression Module, which provides a straightforward strategy to extract the most representative items from these historical interactions, thereby assisting the LLM-driven agent in efficiently handling historical information.
Despite the agent's human-like behavior, its interactions inevitably deviate from authentic user preferences, and these deviations accumulate throughout the interaction process.
To address this potential issue, we design the Preference Stabilization Module, which stabilizes the agent's preferences during early interactions, thereby preventing preference drift in subsequent interactions.
Furthermore, we introduce a debiasing process to mitigate both the exposure bias inherent in the autoregressive framework and the position bias from the LLM-driven agent. 
These improvements enable our generated data to achieve higher item space coverage and better approximate the original training data.

We also assess how recommendation list length, model architecture and LLM affect susceptibility to MEAs, and suggest a simple defense to reduce these risks. 
The aim of this work is to raise awareness of the security and privacy risks of MEAs in recommender systems.

In summary, this work makes the following contributions: 
\begin{itemize}
    \item We identify the pattern mismatch in existing MEA data generation methods and explore LLMs' potential for simulating realistic user-recommender interactions in data generation of MEAs.
    \item We propose an LLM data generation framework with memory compression and preference stabilization, along with debiasing techniques to address exposure and position biases.
    \item Extensive experiments demonstrate that our method significantly improves the quality of generated data and enhances attack performance on sequential recommenders. 
The code is available\footnote{\url{https://github.com/Silung/LLM4MEA}}.
\end{itemize}

\section{Related Work}
\subsection{Model Extraction against RecSys}

Model extraction attacks have been extensively studied in image and text classification but remain underexplored in recommender systems. The study SR-DFME~\cite{yue2021black} was the first to expose recommender systems to such attacks using synthetic data. Subsequent research, such as~\cite{zhu2023model}, has advanced attack methods by combining limited target data with auxiliary data via attention mechanisms and stealing functions to enhance attack performance. Recently, \cite{zhang2024few} proposed a few-shot extraction framework that achieves high surrogate model accuracy with less than 10\% of raw data. However, its reliance on user interaction history limits its applicability in broader attack scenarios. Although the work in \cite{wang2023data} can be viewed as a data-free extraction attack in a white-box setting, it does not fully align with the strict definition of a stealing attack. 

To counter these threats, \cite{zhang2024defense} proposed Gradient-based Ranking Optimization (GRO), a defense strategy that effectively protects recommender systems from model extraction. Despite these advancements, research on both attack and defense strategies for recommender systems remains insufficient, indicating the need for further exploration in this domain.

\subsection{LLMs for Ranking in RecSys}
Large Language Models (LLMs) are transforming ranking methods in recommender systems by enhancing personalization, contextual awareness, and ranking accuracy. 

In \cite{vats2024exploring}, studies on LLMs in recommender systems are categorized by system characteristics and key stages, including LLM-powered \cite{yue2023llamarec, zhang2023collm, wang2023llm4vis, Verma_2023, geng2023recommendation, lei2023recexplainer}, Off-the-shelf \cite{wang2023large, wang2023meditab, du2023enhancing, mysore2023large, wang2023llm4vis}, sequential \cite{ma2024plugin, xiao2024gmeta, liu2024endtoend, petrov2023generative, liao2023llara}, conversational \cite{feng2023large, gao2023chatrec}, personalized \cite{yang2023palr, zhang2023recommendation, li2024paprec}, and knowledge graph-enhanced \cite{wang2024enhancing, xi2023openworld} methods. These approaches enhance recommendation accuracy, personalization, and interpretability.

LLamaRec \cite{yue2023llamarec}, designed for ranking-based recommendations, inspired our work. It retrieves candidates based on user history and uses LLMs for efficient ranking, showcasing the potential of LLMs to improve recommendation performance in diverse applications.

\section{Preliminaries}
\begin{figure*}[]
  \centering
  \includegraphics[width=\linewidth]{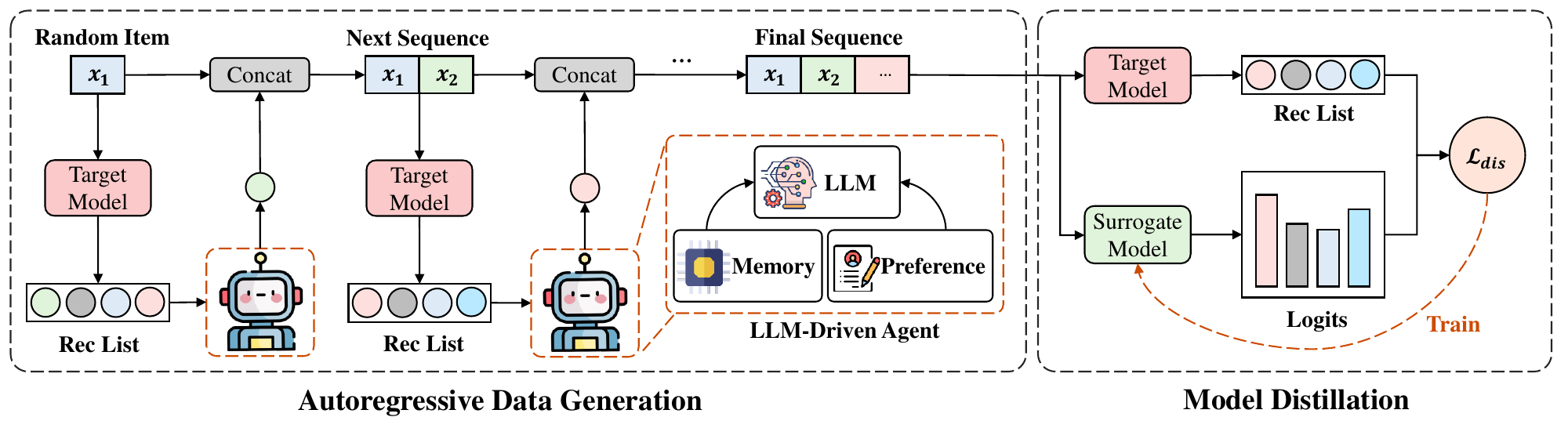}
  \caption{Framework overview. Left: Sequences are autoregressively generated by querying the target model, with item selection guided by an LLM-driven agent employing Memory Compression (MC) and Preference Stabilization (PS) module. Right: The final sequences and target model's outputs are used to train a surrogate model via distillation.}
  \label{fig:ours_framework}
\end{figure*}

\subsection{Threat Model}
\label{threat}
\textbf{Attack Interface:} We consider a black-box setting where the attacker interacts with the target system solely via clicks, without access to model parameters or training data (\textbf{Data-free}). As in~\cite{zhang2024few}, we also evaluate a \textbf{Data-limited} setting with short sequences from few users. Following in~\cite{yue2021black}, we consider scenarios with known model architecture to evaluate MEA performance. \\
\textbf{Item Visibility:} All items and their side information (e.g., titles, categories) are assumed to be visible to the attacker.

\subsection{Model Extraction}
Model Extraction Attacks (MEAs) acquire input-output pairs by querying a recommender system with input sequences and collecting its top-$k$ ranked outputs. These pairs enable training of surrogate models that mimic the target system, supporting unauthorized replication and downstream attacks.

\subsubsection{Secret Data and Target Model} 
Let \(\mathcal{I}\) be the item space with \(|\mathcal{I}|\) items, and \(\mathcal{D}_t\) the secret training data. A black-box target model \(f_t\), trained on \(\mathcal{D}_t\), takes an input sequence \(x = [i_1, i_2, \dots, i_T]\), \(i_j \in \mathcal{I}\), and outputs a next-item distribution \(P(i_{T+1} \mid x)\) over \(\mathcal{I}\). In practice, only a top-\(k\) ranked list \(\hat{L}^k(x) = [r_1, r_2, \dots, r_k]\), \(r_j \in \mathcal{I}\), is returned without probabilities.

\subsubsection{Surrogate Data and Surrogate Model} 
We construct an input sequence set \(\mathcal{X} = \{x_1, x_2, \dots, x_B\}\) via a generation algorithm, where each \(x\) denotes a user interaction sequence. The surrogate dataset \(\mathcal{D}_s\) is obtained by querying the target model:
\begin{equation}
    \mathcal{D}_s = \{(x, \hat{L}^k(x)) \mid x \in \mathcal{X}\}.
\end{equation}

To approximate the target model \(f_t\), we optimize a surrogate model \(f_s\) via ranking-based knowledge distillation:
\begin{equation}
    f^*_{s} = \arg\min_{f_{s}} \sum_{x \in \mathcal{X}} \mathcal{L}_{\mathit{dis}}(f_t(x), f_s(x)),
\end{equation}
where \(\mathcal{L}_{\mathit{dis}}\) quantifies ranking differences.

Following~\cite{yue2021black}, we define \(\hat{S}^k = [f_s(x)_{[\hat{L}^k_{[i]}]}]^k_{i=1}\) as scores from \(f_s\) ordered by \(\hat{L}^k\), and \(\hat{S}_{\mathit{neg}}^k\) as scores of \(n\) uniformly sampled negative items. The distillation loss is:
\begin{equation}
\begin{split}
    \mathcal{L}_{\mathit{dis}} = \frac{1}{k-1} \sum_{i=1}^{k-1} \max(0, \hat{S}^k_{[i+1]} - \hat{S}^k_{[i]} + \lambda_1) \\
    + \frac{1}{k} \sum_{i=1}^{k} \max(0, \hat{S}_{\mathit{neg}[i]}^k - \hat{S}^k_{[i]} + \lambda_2),
\end{split}
\end{equation}
where \(\lambda_1, \lambda_2\) are margin hyperparameters. The first term preserves top-k ranking order; the second penalizes negatives scoring higher than positives.

\section{Method}
\subsection{Overview of LLM4MEA}
LLM4MEA enhances Model Extraction Attacks by leveraging Large Language Models (LLMs) for realistic user behavior simulation. As detailed in~\Cref{sec:ranker}, the LLM-driven agent generates interaction sequences via a Memory Compression (MC) module for historical encoding and a Preference Stabilization (PS) module for inferring stable preferences.
The surrogate model is then trained via distillation to replicate the target model.
To reduce autoregressive and LLM-induced biases, we apply debiasing strategies from~\Cref{sec:debias}, including random sampling for diverse item exposure and list shuffling to counter position bias.
By improving surrogate data quality, LLM4MEA yields stronger extraction performance, contributing to a deeper understanding of vulnerabilities in sequential recommenders and informing future defenses.

\subsection{LLM-driven Agent}
\label{sec:ranker}
\begin{figure}[]
  \centering
  \includegraphics[width=0.95\linewidth]{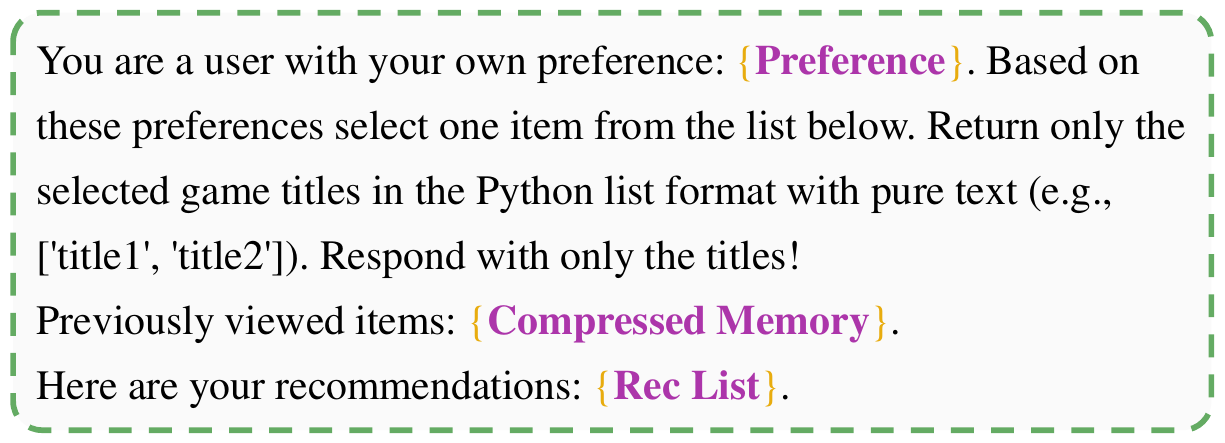}
  \caption{The prompt template integrates outputs from the MC and PS modules, denoted as \textit{Compressed Memory} and \textit{Preference}, respectively. \textit{Rec List} denotes the target model's recommendation.
}
  \label{fig:ranker_prompt}
\end{figure}

The LLM-driven agent simulates user behavior by receiving recommendations from the target model and selecting the next item based on inferred preferences, serving as the \(\text{Sampler}\) in~\Cref{eq:auto}. To adapt LLMs for this task, we introduce the Memory Compression (MC) and Preference Stabilization (PS) modules. The MC module summarizes past interactions into a concise history, while the PS module infers stable preferences to guide decision-making. The LLM, prompted as shown in~\Cref{fig:ranker_prompt}, integrates these inputs with its internal knowledge to select the next item. Module details follow.

\subsubsection{Memory Compression Module}
In sequential recommendation, historical interactions encode essential user preferences and temporal patterns. 
However, as interaction histories grow longer in autoregressive frameworks, they pose challenges for LLMs due to increased computational costs and limited context utilization~\cite{an2024make, li2024long}.

To address this, we propose a Memory Compression (MC) Module that summarizes interaction histories into a fixed-size subset. Given a history sequence \(x\) and a predefined capacity \(\text{size}\), the MC module retains the most representative \(\text{size}\) items when \(|x| > \text{size}\). Motivated by cognitive psychology, we use a heuristic based on the serial position effect~\cite{wong2018serial}, which suggests that the beginning (primacy effect) and end (recency effect) of a sequence are more memorable than the middle.

Guided by this intuition, the MC module retains only the first \(\left\lfloor \text{size}/2 \right\rfloor\) and the last \(\left\lfloor \text{size}/2 \right\rfloor\) items, discarding the middle part. This design balances long-term stable preferences and short-term behavioral shifts. Formally:
\begin{equation}
MC(x) = [i_1, \dots, i_{\lfloor \text{size}/2 \rfloor}, i_{T-\lfloor \text{size}/2 \rfloor+1}, \dots, i_T], \quad \text{if } T > \text{size}
\end{equation}
where \(T\) is the length of sequence \(x\) and \(i \in x\).

This strategy provides an efficient summary of user history, facilitating preference modeling under memory constraints.

\subsubsection{Preference Stabilization Module}
To maintain consistent selection behavior, the LLM-driven agent needs stable preference guidance derived from historical interactions. The Preference Stabilization (PS) module addresses this need by extracting and maintaining a preference profile that guides the agent's decisions.

\begin{figure}[]
    \centering
    \includegraphics[width=0.8\linewidth]{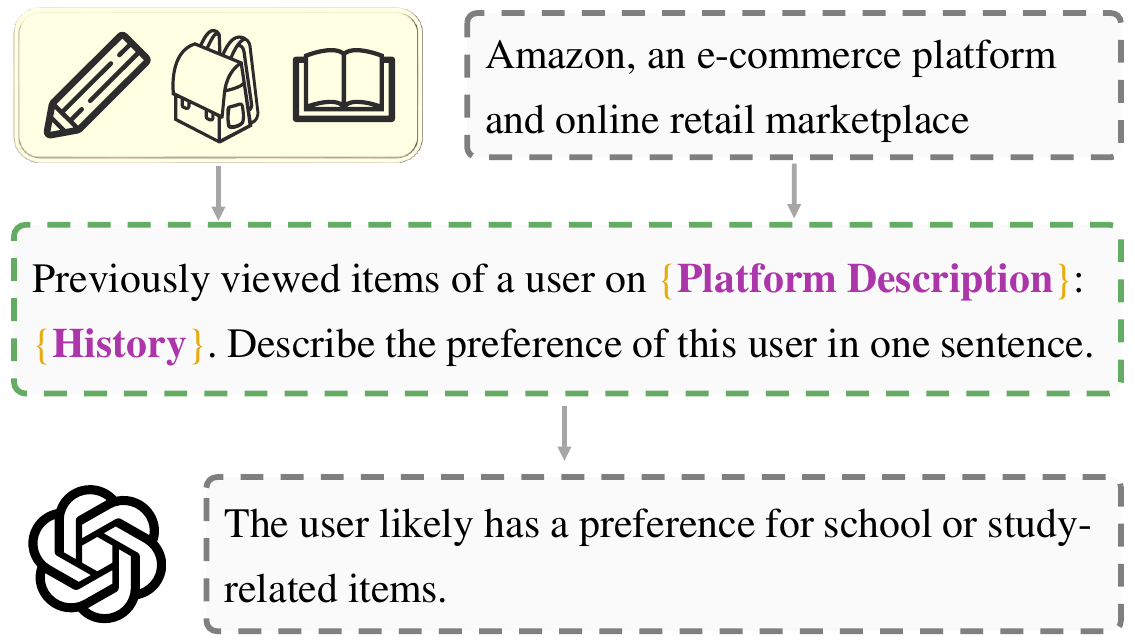}
    \caption{The Preference Stabilization (PS) module summarizes user preferences by prompting the LLM with the \textit{Platform Description} and user \textit{History}.}
  \label{fig:ps_prompt}
\end{figure}

Considering the complexity of user preferences and the limited information in short sequences, the PS module is activated after the interaction history reaches length \(n\).
As shown in~\Cref{fig:ps_prompt}, the LLM processes the historical interactions and generates a preference summary that captures key behavioral patterns, including frequently selected item types and contextual dependencies.
By integrating this stabilized preferences into the decision-making process, the agent can emulate a consistent behavioral pattern, ensuring that its actions align with long-term trends in user behavior rather than being overly influenced by short-term fluctuations.

\subsubsection{Accelerate}
\label{sec:acc}
The autoregressive data generation process is inherently sequential and non-parallelizable, requiring LLM-driven agent at each step to extend the sequence by one item. 
This sequential nature incurs significant computational overhead, particularly for long sequences. 
Moreover, in real-world scenarios, users often show interest in multiple items from a single recommendation list, which contradicts the single-item selection behavior of the current agent.

To mitigate this limitation, we propose an acceleration method based on~\Cref{eq:auto} that enables simultaneous multi-item selection. This enhancement is easily implemented by modifying the internal prompt of the LLM-driven agent:
\begin{align}
    x^{(j+n)} = [x^{(j)}; \text{Agent}(f_t(x^{(j)}))] = [x^{(j)}; i_{j+1}; i_{j+2}; \dots; i_{j+n}]
    \label{eq:auto}
\end{align}
where \(n\) is the number of items selected in each iteration.

This approach not only reduces computational overhead but also better simulates natural user behavior.

\subsection{Debias}
\label{sec:debias}
In training the recommender system, it is crucial to account for the potential biases between the training data and the real-world data. 
Similarly, biases between surrogate and secret data can significantly degrade MEA performance, particularly the exposure bias from the target model and position bias from the agent.
These biases are amplified by repeated interactions in the autoregressive generation process. Mitigating such discrepancies is essential for generating high-quality surrogate data.

\subsubsection{Exposure Bias}
Our autoregressive framework inherits exposure bias from the target recommender, as the agent can only select items from exposed recommendations.
This limits access to the full item space $\mathcal{I}$, 
resulting in biased data that overrepresents frequently recommended items and underrepresents others, potentially distorting surrogate model training.

To address this, we augment the LLM-driven agent with a random sampler that selects items uniformly from $\mathcal{I}$:
\begin{equation}
x = \{i_j \mid i_j \sim \text{Uni}(\mathcal{I}), j = 1, 2, \dots, T.\},
\label{eq:random_sampling}
\end{equation}
where $\text{Uni}(\mathcal{I})$ denotes the uniform distribution over $\mathcal{I}$.

To achieve an expected coverage of \(m\) distinct items from \(|\mathcal{I}|\), we compute the required number of samples \(K\) using the Coupon Collector’s Problem, where the expectation is:
\begin{equation}
    \mathbb{E}[K] = |\mathcal{I}| \sum_{i=|\mathcal{I}|-m+1}^{|\mathcal{I}|} \frac{1}{i},
\end{equation}
We set $m = 0.9|\mathcal{I}|$ to ensure high item coverage.
Combining samples from both the agent and random sampler yields more diverse training data, mitigating exposure bias.

\subsubsection{Position Bias}
Position bias occurs when users interact more with items placed in prominent positions, skewing the representation of user preferences. In our framework, the LLM-driven agent exhibits this bias, exacerbated by the autoregressive setup. To mitigate this, we shuffle the target model’s recommendation list before presenting it to the agent, ensuring random interaction order and preventing position bias. This reduces the distortion caused by position bias, improving the quality and reliability of the data for training surrogate models.

\section{Experiments}
In this section, we try to address the following research questions:
\textbf{RQ1}: How well does LLM4MEA perform in model extraction? \\
\textbf{RQ2}: Does our method generate more realistic data compared to existing methods? \\
\textbf{RQ3}: How does exposure bias and position bias affect the data generation process? \\
\textbf{RQ4}: What additional factors influence MEA performance? \\
\textbf{RQ5}: How to defend recommender systems from MEA?\\

\subsection{Setup}
\subsubsection{Dataset}
We use three well-known recommendation datasets listed in~\Cref{tab:dataset_statistics}: Beauty and Games datasets from the Amazon product reviews~\cite{he2016ups,mcauley2015image}, and Steam dataset from the Steam video game platform~\cite{kang2018self,wan2018item,pathak2017generating}. 
The data is processed into implicit feedback based on the rating information. 
Similar to prior studies~\cite{yue2021black,sun2019bert4rec}, the last two items in each user sequence are held out for validation and testing, while the rest are used for training. The black-box target model is set to return the top-100 recommended items for each query.

\begin{table}[]
\centering
\caption{Dataset Statistics}
\label{tab:dataset_statistics}
\resizebox{\linewidth}{!}{
\begin{tabular}{lrrrrrr}
\toprule
\textbf{Datasets} & \textbf{Users} & \textbf{Items} & \textbf{Interactions} & \textbf{Avg. len} & \textbf{Max. len} & \textbf{Sparsity} \\ 
\midrule
Beauty & 40,226 & 54,542 & 353,962 & 8.79 & 291 & 99.98\% \\ 
Games  & 29,341 & 23,464 & 280,945 & 9.57 & 858 & 99.95\% \\ 
Steam  & 334,542  & 13,046 & 4,212,380 & 12.59 & 2,043 & 99.90\% \\ 
\bottomrule
\end{tabular}
}
\end{table}

\subsubsection{Target Model Arch}
To evaluate the performance of attacks, we implement the model extraction attack on three representative sequential recommenders: NARM \cite{li2017neural}, SASRec \cite{kang2018self}, and BERT4Rec \cite{sun2019bert4rec}. 
These models differ in their architectural components and training schemes.

\subsubsection{Baselines}
We compare our method against two baselines: Random and SR-DFME \cite{yue2021black}. 

\textbf{Random} method is the simplest approach,, where each item in the sequence is independently selected from the item space \( \mathcal{I} \), without any contextual dependency. 
Formally, for a sequence \( x = \{i_1, i_2, \dots, i_T\} \in \mathcal{X} \) is generated such that each \( i \in \mathcal{I} \) is chosen independently, as described in~\Cref{eq:random_sampling}.

\textbf{SR-DFME} (\textbf{S}equential \textbf{R}ecommenders via \textbf{D}ata-\textbf{F}ree \textbf{M}odel \textbf{E}xtraction) employs an autoregressive process with a random sampler, where each item in the sequence is predicted based on the previously selected items. 

\subsubsection{Metrics}
Following~\cite{yue2021black}, we evaluate the attack performance using both recommendation and extraction metrics. 
Additionally, a data fidelity metric is proposed to mesure the quality of surrogate data.

\textbf{Recommendation Metrics}: We use Truncated Recall@K (equivalent to HR@K) and NDCG@K to assess ranking quality. For efficiency, we sample 100 negative items uniformly for each user and evaluate them with the positive item.

\textbf{Extraction Metrics}: Agreement@K measures the output similarity between two models:
    \begin{equation}
        \mathit{Agreement}@K = {|\mathit{B}_\text{topK}\cap \mathit{W}_\text{topK}|}/{K},
    \end{equation}
    where \( \mathit{B}_\text{topK} \) is the top-K recommendation list from the black-box target model \(f_t\) and \( \mathit{W}_\text{topK} \) is from our white-box surrogate model \(f_s\).

\textbf{Data Fidelity Metric}:\label{subsection:n-gram} 
    To quantify the fidelity of generated data, we propose N-gram Div metrics that measure the KL-divergence between secret and surrogate data's N-gram distributions:
    \begin{equation}
        N\text{-gram Div} = \sum_{g} P_\epsilon(g) \log[ {P_\epsilon(g)}/{Q_\epsilon(g)}],
    \end{equation}
    where \(g\) represents the n-gram terms, \(P_\epsilon(g)\) and \(Q_\epsilon(g)\) are adjusted N-gram probabilities with smoothing factor \(\epsilon\) for zero probabilities.

We report results for \(K = 1, 10\) in Agreement@K and \(N = 1, 2\) in N-gram Div.

\begin{table*}[]
\centering
\caption{Extraction performance under identical model architecture and 5k sequences.}
\label{tab:main_result}
\resizebox{\textwidth}{!}{
\begin{tabular}{ccccccccccccccc}
\toprule
\multirow{2}{*}{Archs} & \multirow{2}{*}{Threats} & \multirow{2}{*}{Method} & \multicolumn{4}{c}{Beauty}   & \multicolumn{4}{c}{Games} & \multicolumn{4}{c}{Steam}  \\ 
 \cmidrule(l){4-7} \cmidrule(l){8-11} \cmidrule(l){12-15}
                &&&N@10 & R@10 & \textbf{Agr@1} & \textbf{Agr@10} & N@10 & R@10 & \textbf{Agr@1} & \textbf{Agr@10} & N@10 & R@10 & \textbf{Agr@1} & \textbf{Agr@10}\\ \midrule
\multirow{11}{*}{NARM} & \multirow{2}{*}{Available} & Target  & 0.3564 & 0.5179 & -  & -  & 0.5452 & 0.7423 & -  & -  & 0.6254 & 0.8465 & -  & -  \\  
& & Secret & 0.3228 & 0.4613 & 0.4765 & 0.6049 & 0.5426 & 0.7397 & 0.4647 & 0.6264 & 0.6263 & 0.8423 & 0.7515 & 0.7655  \\ \cmidrule(l){2-15}
& \multirow{5}{*}{Free} & Random & 0.3167 & 0.4704 & 0.4386 & \underline{0.6258} & 0.5203 & 0.7218 & 0.4041 & 0.5761 & 0.6262 & 0.8484 & 0.6685 & 0.6901 \\  
& & SR-DFME  & 0.2710 & 0.3641 & 0.2733 & 0.3209 & 0.5358 & 0.7273 & 0.4395 & 0.6066 & 0.6165 & 0.8303 & 0.7483 & 0.7478  \\  
& &Ours-eBias  & 0.2548 & 0.3618 & 0.2907 & 0.3996 & 0.5378 & 0.7293 & 0.4580 & 0.6215 & 0.6160 & 0.8306 & 0.7523 & 0.7457  \\
& & Ours-pBias & 0.3181 & 0.4636 & \underline{0.4450} & 0.5980 & 0.5347 & 0.7285 & \underline{0.4916} & \underline{0.6659} & 0.6212 & 0.8401 & \underline{0.7675} & \underline{0.7678}\\
& & Ours       & 0.3221 & 0.4672 & \textbf{0.4957} & \textbf{0.6393} & 0.5341 & 0.7275 & \textbf{0.4998} & \textbf{0.6734} & 0.6248 & 0.8431 & \textbf{0.7723} & \textbf{0.7812} \\ \cmidrule(l){2-15}
& \multirow{4}{*}{Limited} & SR-DFME  & 0.2876 & 0.4124 & 0.3477 & 0.4656 & 0.5395 & 0.7319 & 0.4473 & 0.6157 & 0.6159 & 0.8283 & 0.7474 & 0.7548  \\  
& & Ours-eBias  & 0.2888 & 0.4110 & 0.3761 & 0.4873 & 0.5408 & 0.7326 & 0.4585 & 0.6279 & 0.6154 & 0.8265 & 0.7509 & 0.7581  \\  
& & Ours-pBias  & 0.3210 & 0.4633 & \textbf{0.4992} & \textbf{0.6427} & 0.5342 & 0.7295 & \underline{0.4917} & \underline{0.6721} & 0.6243 & 0.8424 & \underline{0.7730} & \textbf{0.7808} \\  
& & Ours  &  0.3108 & 0.4505 & \underline{0.4468} & \underline{0.5919} & 0.5378 & 0.7327 & \textbf{0.5073} & \textbf{0.6906} & 0.6284 & 0.8487 & \textbf{0.7764} & \underline{0.7763} \\ \midrule 
\multirow{11}{*}{SASRec} & \multirow{2}{*}{Available} & Target  & 0.2558 & 0.4412 & -  & -  & 0.3606 & 0.5878 & -  & -  & 0.6110 & 0.8368 & -  & -  \\  
& & Secret  & 0.2260 & 0.3720 & 0.6330 & 0.7528 & 0.3492 & 0.5609 & 0.6360 & 0.7497 & 0.6127 & 0.8344 & 0.7556 & 0.8128  \\  \cmidrule(l){2-15}
& \multirow{5}{*}{Free} & Random & 0.2259 & 0.3808 & 0.3829 & 0.6414 & 0.2908 & 0.4775 & 0.3639 & 0.5415 & 0.6174 & 0.8416 & 0.6490 & 0.7204 \\
& & SR-DFME  & 0.1302 & 0.1973 & 0.3189 & 0.5969 & 0.2969 & 0.4573 & 0.4807 & 0.6080 & 0.6112 & 0.8330 & 0.6999 & 0.7625  \\  
& & Ours-eBias  & 0.1321 & 0.1949 & 0.4014 & 0.6215 & 0.3002 & 0.4650 & 0.5057 & 0.6315 & 0.6095 & 0.8311 & 0.7339 & 0.7910  \\ 
& & Ours-pBias & 0.2020 & 0.3327 & \underline{0.4962} & \underline{0.6622} & 0.3305 & 0.5302 & \underline{0.5886} & \underline{0.7008} & 0.6101 & 0.8319 & \underline{0.7486} & \underline{0.8035} \\ 
& & Ours       & 0.2131 & 0.3538 & \textbf{0.5785} & \textbf{0.7200} & 0.3321 & 0.5318 & \textbf{0.5995} & \textbf{0.7121} & 0.6038 & 0.8202 & \textbf{0.7505} & \textbf{0.8073}  \\ \cmidrule(l){2-15}
& \multirow{4}{*}{Limited} & SR-DFME  & 0.1610 & 0.2456 & 0.5144 & \underline{0.6745} & 0.3204 & 0.5045 & 0.5169 & 0.6577 & 0.6032 & 0.8200 & 0.6755 & 0.7483  \\  
& & Ours-eBias  & 0.1654 & 0.2524 & \underline{0.5298} & 0.6565 & 0.3315 & 0.5236 & 0.5399 & 0.6548 & 0.6040 & 0.8218 & 0.7093 & 0.7668  \\  
& & Ours-pBias  & 0.2136 & 0.3546 & \textbf{0.5720} & \textbf{0.7130} & 0.3394 & 0.5448 & \underline{0.6141} & \underline{0.7098} & 0.6033 & 0.8200 & \textbf{0.7506} & \underline{0.8028} \\ 
& & Ours  & 0.2057 & 0.3385 & 0.4680 & 0.6349 & 0.3259 & 0.5247 & \textbf{0.6219} & \textbf{0.7264} & 0.6156 & 0.8394 & \underline{0.7501} & \textbf{0.8161} \\ \midrule 
\multirow{11}{*}{BERT4Rec}  & \multirow{2}{*}{Available} & Target  & 0.3212 & 0.4937 & -  & -  & 0.3890 & 0.6124 & -  & -  & 0.6178 & 0.8451 & -  & -  \\  
& & Secret  & 0.1990 & 0.2886 & 0.6773 & 0.7053 & 0.2825 & 0.4218 & 0.5972 & 0.6346 & 0.6197 & 0.8414 & 0.7845 & 0.7735  \\  \cmidrule(l){2-15}
& \multirow{5}{*}{Free} & Random & 0.1976 & 0.2911 & \textbf{0.6541} & \textbf{0.7009} & 0.3093 & 0.4654 & 0.3761 & \textbf{0.5636} & 0.6226 & 0.8467 & 0.5027 & 0.5843 \\  
& & SR-DFME  & 0.1395 & 0.2084 & 0.1827 & 0.2439 & 0.2456 & 0.3422 & \underline{0.4608} & 0.4672 & 0.6102 & 0.8309 & 0.7514 & 0.7293  \\  
& & Ours-eBias  & 0.1504 & 0.2224 & 0.1999 & 0.2856 & 0.2527 & 0.3516 & \textbf{0.5138} & \underline{0.5321} & 0.6075 & 0.8253 & 0.7516 & 0.7308  \\ 
& & Ours-pBias & 0.1894 & 0.2725 & 0.5812 & 0.6299 & 0.2552 & 0.3578 & 0.4010 & 0.4685 & 0.6129 & 0.8337 & \textbf{0.7705} & \textbf{0.7541} \\ 
& & Ours       & 0.1832 & 0.2626 & \underline{0.6299} & \underline{0.6737} & 0.2636 & 0.3705 & 0.4205 & 0.5123 & 0.6120 & 0.8330 & \underline{0.7684} & \underline{0.7535} \\ \cmidrule(l){2-15}
& \multirow{4}{*}{Limited} & SR-DFME  & 0.1634 & 0.2291 & 0.5926 & 0.6137 & 0.2642 & 0.3757 & \underline{0.6046} & 0.6146  & 0.6065 & 0.8264 & 0.7414 & 0.7177  \\  
& & Ours-eBias  & 0.1647 & 0.2315 & \underline{0.6188} & \underline{0.6400} & 0.2668 & 0.3787 & \textbf{0.6204} & \underline{0.6218} & 0.6080 & 0.8271 & 0.7538 & 0.7409  \\ 
& & Ours-pBias  &  0.1891 & 0.2758 & \textbf{0.6549} & \textbf{0.7011} & 0.2762 & 0.3970 & 0.5553 & 0.5968 & 0.6154 & 0.8355 & \underline{0.7746} & \textbf{0.7630} \\ 
& & Ours  & 0.1835 & 0.2634 & 0.5726 & 0.6169 & 0.2965 & 0.4348 & 0.5657 & \textbf{0.6474} & 0.6165 & 0.8396 & \textbf{0.7783} & \underline{0.7509} \\ \bottomrule 
\end{tabular}
}
\end{table*}

\subsection{Attack Performance (RQ1)}
\subsubsection{Identical Arch}
We evaluate model extraction attacks under three data access scenarios: Data-free (Free), and Data-limited (Limited), as described in~\Cref{threat}. 
Additionally, we include a Data-available (Available) setting with full access to secret data as a reference, where we evaluate both the target model (\textbf{Target}) and a surrogate model trained on a size-constrained subset of secret data (\textbf{Secret}).

For synthetic data generation, we compare three methods: random generation (\textbf{Random}), SR-DFME~\cite{yue2021black}, and our proposed approach (\textbf{Ours}). 
We also analyze two variants of our approach: \textbf{Ours-eBias} which is affected by exposure bias and \textbf{Ours-pBias} which is affected by position bias, as detailed in~\Cref{section:ebias} and~\Cref{section:pbias}, respectively.
Results are shown in~\Cref{tab:main_result}).

\textbf{Observation.}
In summary, 
our method with debiasing achieves superior MEA performance while maintaining recommendation accuracy. In the data-free setting, it outperforms SR-DFME by 51.90\% in Agreement@1 and 37.74\% in Agreement@10, and surpasses Random by 22.24\% and 10.99\% respectively. 
The data-limited setting shows more consistent improvements, with 7.03\% higher Agreement@1 and 7.36\% higher Agreement@10 compared to SR-DFME, along with reduced performance variance across scenarios.

Our method outperforms baselines in extracting NARM and SASRec across both data-free and data-limited settings. For BERT4Rec, while achieving the best results on Steam, performance on Beauty is sub-optimal, likely due to high sparsity (99.98\%) and weak sequential patterns. 

We also observe that models trained on dense datasets like Steam are more vulnerable to extraction attacks. Notably, the extracted models maintain high performance on the original recommendation task.

\subsubsection{Cross Arch}
\begin{figure}[]
    \centering
    \includegraphics[width=1\linewidth]{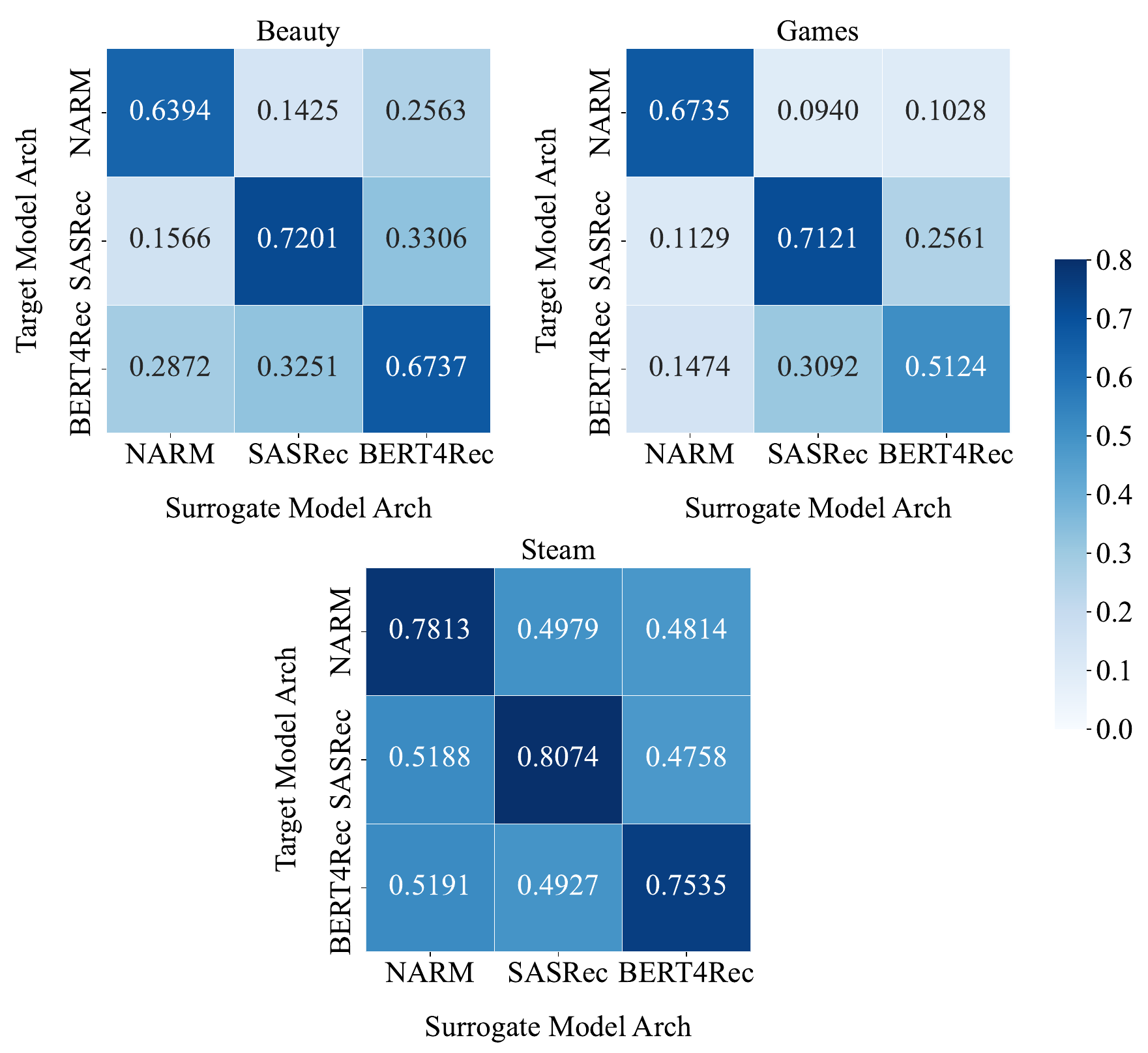}
    \caption{Agr@10 of Model Extraction Attack among different architectures.}
    \label{fig:cross}
\end{figure}

We investigate how MEA performs when the surrogate model architecture differs from the target model. 
Results are presented as heatmaps in~\Cref{fig:cross}, where horizontal and vertical axes denote surrogate and target architectures, respectively.

\textbf{Observation.}
MEA achieves optimal performance when the surrogate architecture matches the target architecture, with performance degrading for mismatched architectures. 
This finding emphasizes the importance of architecture confidentiality in protecting model privacy. 
Additionally, MEA demonstrates superior performance on dense datasets (Steam) compared to sparse ones, consistent with previous study~\cite{yue2021black}.

\subsection{Data Divergence (RQ2)}
\label{sec:div}
Since secret datasets represent real user behaviors, we consider surrogate data more authentic when it exhibits smaller divergence from secret datasets.
To quantify this divergence, we utilize the N-gram Div metric described in~\Cref{subsection:n-gram}, measuring the divergence between surrogate and secret datasets.
The N-gram Div is calculated using uni-gram and bi-gram (i.e., \(N=1,2\)) across various datasets and model architectures.
All experiments use identical model architectures and are conducted with 5k sequences.

\begin{figure}[]
      \centering
      \hspace*{-0.6cm}
      \includegraphics[width=1.12\linewidth]{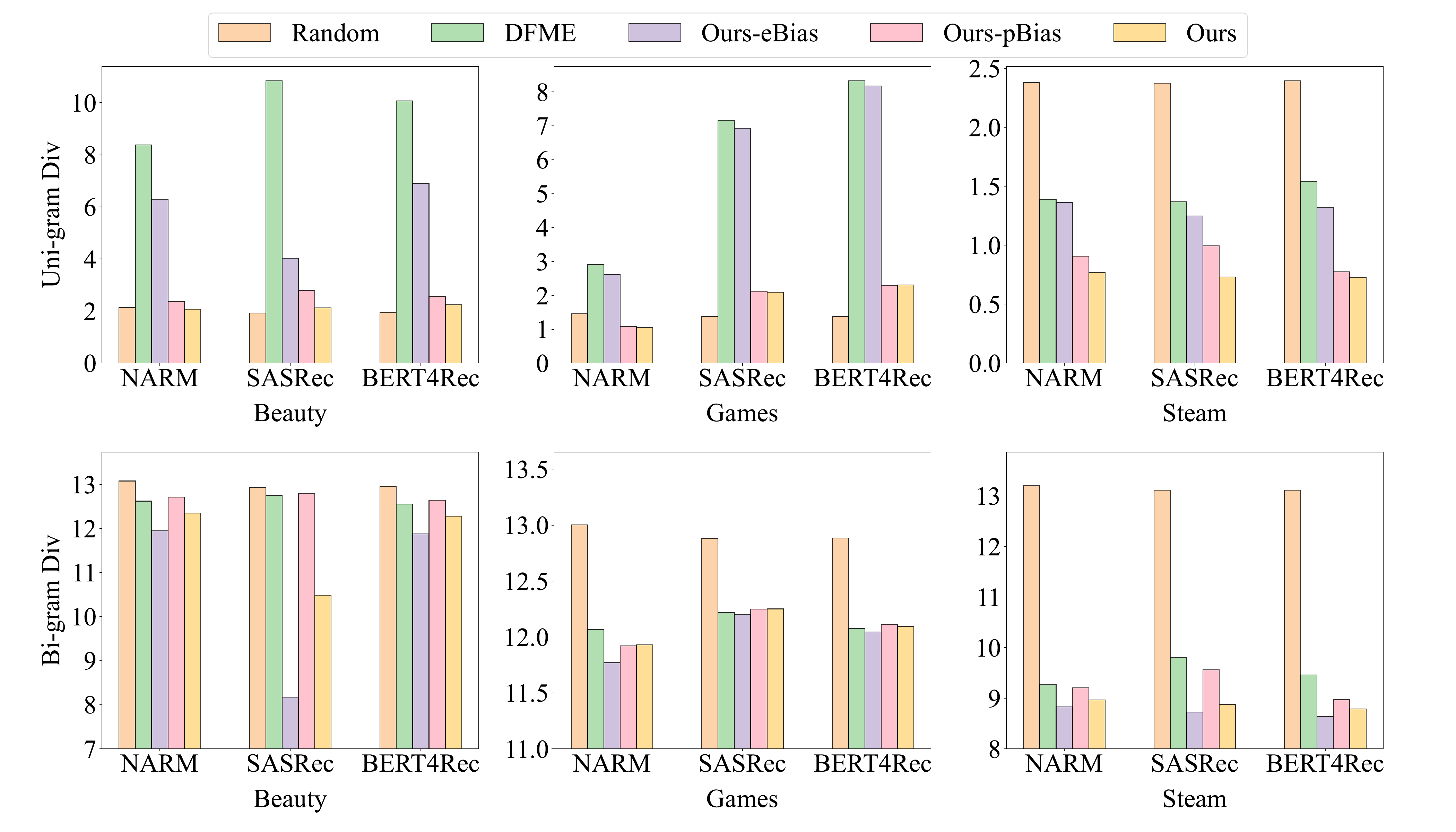}
  \caption{The Uni-gram Div and Bi-gram Div metric between surrogate and secret data in the data-free setting. Lower value means the gap between surrogate and secret data is smaller.}
  \label{fig:ngram_div}
\end{figure}

The results are shown in~\Cref{fig:ngram_div}, where the y-axis represents the N-gram Div value. A lower value indicates smaller divergence from secret data and thus higher authenticity of the generated data.
The N-gram Div evaluates data authenticity from two perspectives: uni-gram (N=1) assesses individual item distributions, while bi-gram (N=2) captures sequential patterns between adjacent items.

\textbf{Observation.}
Our analysis shows that our method significantly reduces the divergence from real data patterns: compared to SR-DFME, it achieves 64.98\% and 4.74\% lower divergence on average for uni-grams and bi-grams respectively; compared to random sampling, it shows 10.67\% and 16.27\% lower divergence on average respectively. 
It's worth noting that in specific cases (e.g., Beauty dataset with BERT4Rec), while our method shows slightly higher Uni-gram Div than Random, it achieves lower Bi-gram Div, indicating better capture of sequential patterns despite minor differences in individual item distributions.

The relative performance ranking of different data generation methods (Random, SR-DFME, Ours and so on.) remains consistent across different target model architectures when extracting the same dataset. 
For instance, when extracting models trained on Steam, our method consistently performs best, followed by Ours-pBias, Ours-eBias, SR-DFME, while Random performs worst, regardless of the target model architecture. 
This consistency suggests that the effectiveness of data generation algorithms \(G\) is relatively independent of the target model architecture. Therefore, when the surrogate model shares the same architecture as the target model, the choice of data generation method for MEA should primarily focus on dataset.

\subsection{The Impact of Biases (RQ3)}
\subsubsection{Exposure Bias from Target Model}
\label{section:ebias}
An ideal data generation algorithm \(G\) should reconstruct patterns in secret data. However, in the autoregressive framework, generated items are restricted to those recommended by the target model. Due to exposure bias, many items rarely or never appear in recommendations, limiting surrogate data diversity and creating a significant gap from secret data.

\begin{figure}[]
    \centering
    \begin{subfigure}[b]{0.52\linewidth}
        \centering
        \includegraphics[width=\linewidth]{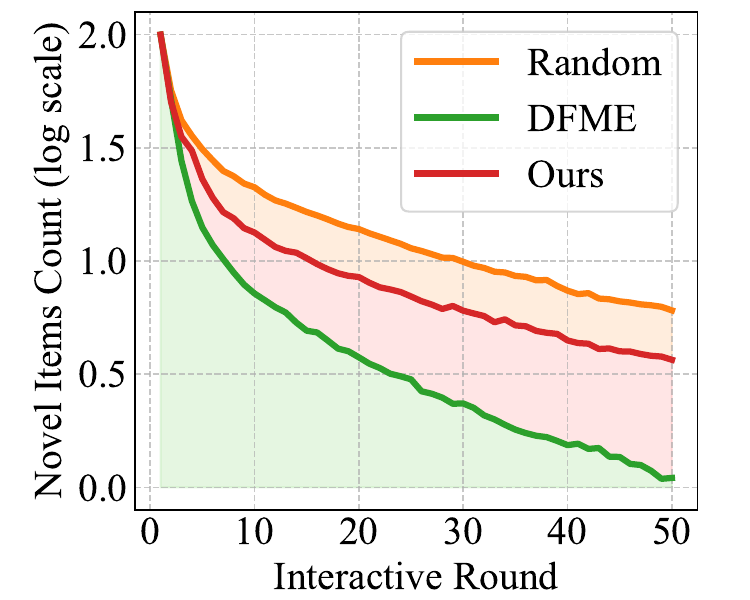}
        \caption{Unseen item counts (log scale) over interactions. The rapid decrease indicates a strong exposure bias. Shaded areas show cumulative exposure differences.}
        \label{fig:exposure_bias}
    \end{subfigure}
    \hfill
    \begin{subfigure}[b]{0.44\linewidth}
        \centering
        \includegraphics[width=\linewidth]{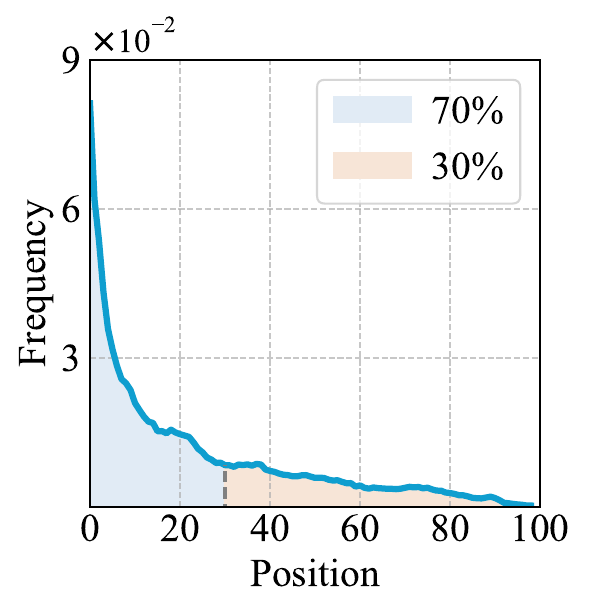}
        \caption{Selection frequency across positions from 100 independent items, showing the LLM's position bias through non-uniform distribution.}
        \label{fig:position_bias}
    \end{subfigure}
    \caption{Analysis of biases: (a) exposure bias in data generation process; (b) position bias in LLM selections.}
    \label{fig:bias_analysis}
\end{figure}

To demonstrate this, we conduct an experiment on the Beauty dataset with NARM architecture. We track the number of unseen items (never previously recommended) as interactions progress. In~\Cref{fig:exposure_bias}, the x-axis represents the interaction round, and the y-axis shows the logarithm (base 10) of unseen item count. The shaded areas between curves indicate gaps in unseen item exposure: the area between Random and Ours reflects additional exposure in Random's generation, while the area between Ours and SR-DFME shows our method's advantage in item coverage. The Random method, selecting items independently of recommendations, serves as a baseline unaffected by exposure bias.

\textbf{Observation.} 
The results reveal that unseen items in the target model's recommendations decrease rapidly as rounds progress, primarily due to the accumulation of previously recommended items. SR-DFME suffers the most severe exposure bias, with almost no unseen items after 50 rounds. Our method significantly mitigates this issue.

In~\Cref{tab:main_result}, comparison between Ours-eBias and Ours shows that mitigating exposure bias significantly enhances MEA performance in the data-free setting, improving Agreement@1 and Agreement@10 by 38.49\% and 26.56\%, respectively. 
In the data-limited setting, improvements are modest (3.23\% and 5.53\% on average) due to two factors: first, the data generated by the random sampler replaces a portion of the secret data; second, the secret data itself is not affected by the exposure bias inherent in the autoregressive framework. These findings suggest that debiasing partially offsets the benefits of secret data.

\subsubsection{Position Bias of LLM-driven Agent}
\label{section:pbias}
The LLM-driven agent in our framework exhibits a position bias, favoring items in certain positions of the recommendation list, similar to human biases. We analyze the position distribution of selections when the agent is presented with 100 randomly chosen, independent items from the Beauty dataset.

\textbf{Observation.}
As shown in~\Cref{fig:position_bias}, the x-axis represents the position index (1-100), and the y-axis shows selection frequency. A long-tail distribution is observed, with higher-positioned items selected more frequently. Specifically, 70\% of selections come from the top 30 positions.
This strong positional preference, independent of item content, demonstrates the inherent position bias of the LLM agent.

Comparing Ours-pBias and Ours in~\Cref{tab:main_result}, avoiding position bias improves MEA's data-free performance by 5.03\% in Agreement@1 and 4.09\% in Agreement@10. However, debiasing has little effect when attacking BERT4Rec on the Steam dataset, and does not enhance MEA performance in the data-limited setting, as the secret data is unaffected by the LLM's position bias.

\subsection{Additional Factors Affecting MEA Performance (RQ4)}
\subsubsection{Using Different LLM to Drive Agent}
We adopt GPT-4o-mini to drive the agent due to its strong balance between efficiency and intelligence. To assess the effect of LLM choice, we evaluate Llama-3-8B-Instruct (Meta), Phi-3-Small-8k-Instruct
% \footnote{\url{https://huggingface.co/microsoft/Phi-3-small-8k-instruct}}
(Microsoft), and Mistral-7B-Instruct-v0.3
% \footnote{\url{https://huggingface.co/mistralai/Mistral-7B-Instruct-v0.3}} 
(Mistral AI), using the same architecture (NARM) and dataset (Beauty). As shown in~\Cref{tab:llms}, GPT-4o-mini yields the best results, though performance differences are marginal across LLMs.

\begin{table}[]
\centering
\caption{Extraction performance using different LLMs under identical model architecture and 5k sequences.}
\label{tab:llms}
\resizebox{\linewidth}{!}{
\begin{tabular}{lccccc}
\toprule
LLM & Params & N@10 & R@10 & Agr@1 & Agr @10 \\ \hline
GPT-4o-mini & - & 0.3221 & 0.4672 & \textbf{0.4957} & \textbf{0.6393} \\
Llama-3-8B-Instruct & 8B & 0.3166 & 0.4612 & 0.4482 & 0.6032 \\
Phi-3-Small-8k-Instruct & 7B & 0.3178 & 0.4623 & 0.4694 & 0.6176 \\
Mistral-7B-Instruct-v0.3 & 7B & 0.3166 & 0.4586 & 0.4831 & 0.6337\\
\bottomrule 
\end{tabular}}
\end{table}

\subsubsection{Recommendation List Length}
\begin{figure}[]
  \centering
  \includegraphics[width=0.7\linewidth]{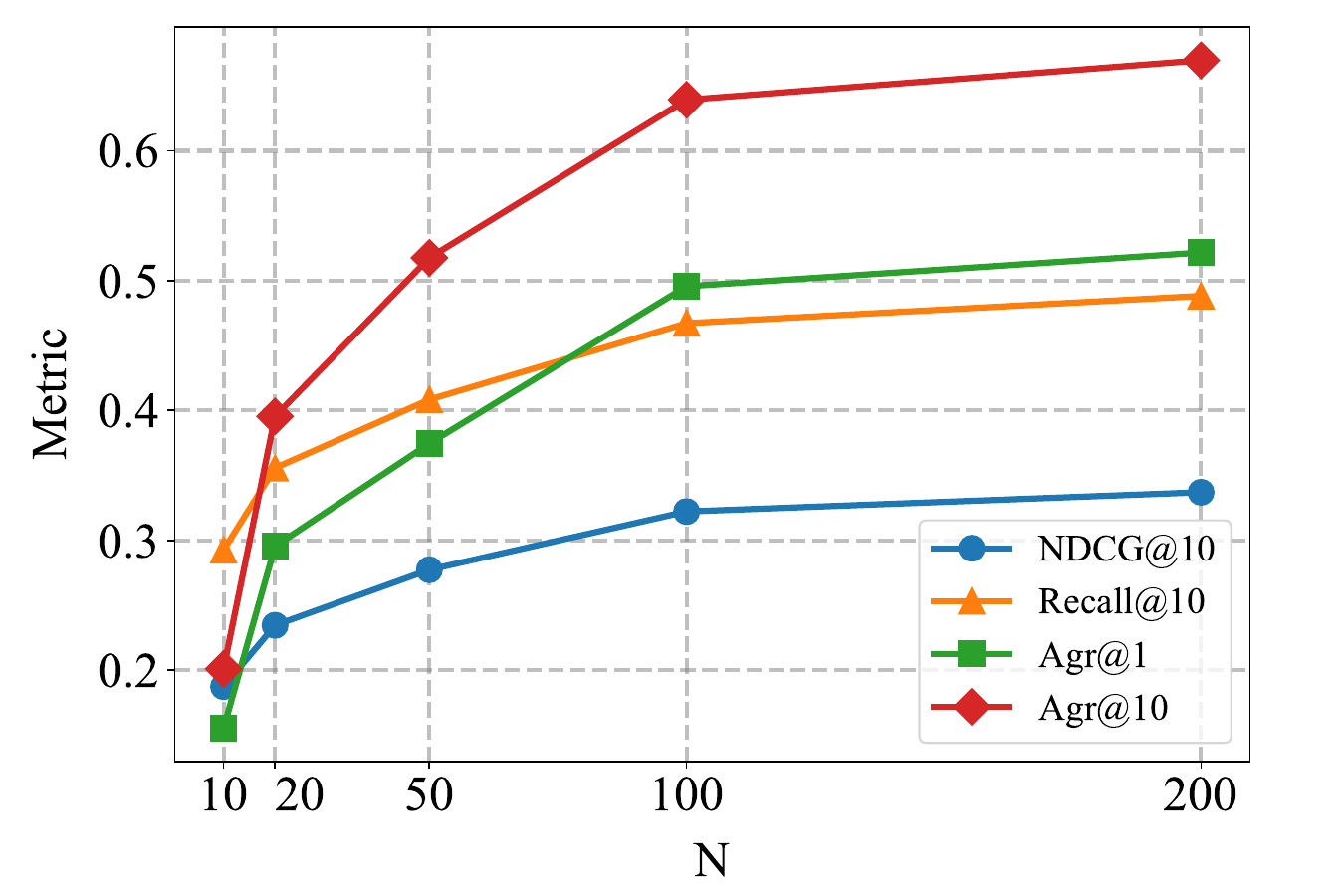}
  \caption{Performance of MEA as the length of the recommendation list \(n\) varies. The results demonstrate that increasing \(n\) enhances the effectiveness of the attack.}
  \label{fig:k}
\end{figure}

A longer recommendation list reveals more information, increasing vulnerability to MEA. We experiment with list lengths \(n\) from 10 to 200, using the NARM architecture on the Beauty dataset. As shown in~\Cref{fig:k}, both the surrogate model's performance and its alignment with the target model improve as the list length grows, highlighting the heightened susceptibility to MEA. From a defense perspective, controlling the recommendation list length offers a practical countermeasure against potential MEA attacks, as users typically do not browse such extensive lists.

\subsubsection{LLM Overhead}
Introducing LLMs incurs manageable computational costs, which can be reduced using acceleration methods in~\Cref{sec:acc}. For example, attacking the NARM model on the Beauty dataset with 5-items/query consumes 102M tokens via the Batch API, taking under 2 hours at a cost of \$7.69. Selecting a single item per query raises the cost to approximately \$38, still a reasonable investment for data collection and training high-performance recommenders. Model Distillation can proceed in parallel while awaiting LLM responses.

\subsection{Defense Strategy (RQ5)}
To provide a balanced view, we explore a lightweight defense against model extraction. For each top-$k$ list, a proportion $p=0.1k$ of items is randomly replaced with items from the item pool, injecting noise into the feedback signal. 
This operation has a slight impact on the recommendation system's performance. 
As shown in~\Cref{tab:defence}, this strategy significantly degrades MEA performance on the Beauty dataset with NARM, validating its effectiveness.

\begin{table}[]
\centering
\caption{MEA performance under defense strategies.}
\small
\begin{tabular}{cccccccc}
\toprule
\textbf{Method} & \textbf{Defense} & \textbf{N@10} & \textbf{R@10} & \textbf{Agr@1} & \textbf{Agr@10} \\
\midrule
\multirow{2}{*}{Random} & \XSolidBrush & 0.3167 & 0.4704 & 0.4386 & 0.6258 &\\
& \Checkmark & 0.2533 & 0.3653 & 0.2624 & 0.3564 & \\  \midrule
\multirow{2}{*}{SR-DFME} & \XSolidBrush & 0.2710 & 0.3641 & 0.2733 & 0.3209 &\\
& \Checkmark & 0.2682 & 0.3598 & 0.2658 & 0.3227 &\\  \midrule
\multirow{2}{*}{LLM4MEA} & \XSolidBrush & 0.3221 & 0.4672 & 0.4957 & 0.6393 &\\
& \Checkmark & 0.2930 & 0.3973 & 0.4290  & 0.4481 &\\
\bottomrule
\end{tabular}
\label{tab:defence}
\end{table}

\section{Conclusion}
In this work, we investigate the vulnerability of sequential recommenders to Model Extraction Attacks and propose the LLM4MEA leveraging LLMs-driven agent to generate data.
Our method addresses the key challenge of generating high-quality surrogate data by simulating realistic user-recommender interactions. 
Through memory compression and preference stabilization modules, along with debiasing techniques, our framework effectively mitigates common issues in synthetic data generation.
Extensive experiments demonstrate that our method significantly outperforms existing approaches in both data quality and attack performance.
The generated data shows higher item space coverage and better approximates real user behavior patterns. 
Our findings highlight the potential of LLMs in improving MEA effectiveness while revealing important insights about recommender systems' vulnerabilities.

Future work could explore the generalization of our framework to other recommendation scenarios and investigate additional defense mechanisms against such attacks.
Understanding these vulnerabilities is crucial for developing more robust recommender systems in an increasingly security-conscious environment.

\section*{Ethics Statement}
This work investigates the vulnerability of sequential recommender systems to MEAs, emphasizing the privacy and security risks associated with unauthorized model replication. Given the pervasive use of recommender systems, our objective is to raise awareness of these threats. 
We propose a simple yet effective defense strategy and identify key hyperparameters of recommender systems that can mitigate the risk of MEAs.
While no definitive solution to MEAs currently exists, our goal is to advance efforts in securing recommender systems, not to facilitate misuse.

% \begin{acks}
% This work is funded by XX, YY, and ZZ. 
% \end{acks}

\bibliographystyle{ACM-Reference-Format}
\bibliography{sample-base}

\appendix
\section{Proof for a Variant of the Coupon Collector Problem}
\begin{proof}
Let \( K_i \) denote the number of samples needed to obtain a new item when there are exactly \( i-1 \) distinct items already sampled. Then, 
\[ K = \sum_{i=n}^{m} K_i. \]

The probability of sampling a new item when there are exactly \( i-1 \) distinct items is given by:  
\[ p_i = 1 - \frac{i-1}{m}. \]

The random variable \( K_i \) follows a geometric distribution with:  
\[
\Pr[K_i = k] = (1 - p_i)^{k-1} p_i,
\]
and the expected value of \( K_i \) is:  
\[
\mathbb{E}[K_i] = \frac{1}{p_i} = \frac{m}{m - i + 1}.
\]

The expected value of \( K \) can then be computed as:  
\[
\mathbb{E}[K] = \mathbb{E}\left[\sum_{i=n}^{m} K_i\right] = \sum_{i=n}^{m} \mathbb{E}[K_i] = \sum_{i=n}^{m} \frac{m}{m - i + 1} = m \sum_{i=m-n+1}^{m} \frac{1}{i}.
\]
\end{proof}

\section{Implementation Details}
We follow the setup in~\cite{yue2021black} with necessary modifications. 
% For each user sequence of length \(T\), we use the first \(T-2\) items for training, leaving the last two for validation and testing. 
Models are trained using Adam optimizer (learning rate: 0.001, weight decay: 0.01) with 100 warmup steps. 
The sequence length is set to 50 with 300 training epochs. 
Batch sizes for NARM, SASRec, and BERT4Rec are 1024, 512, and 512, respectively. 
We primarily use GPT-4o mini (2024-07-18) to drive our agent and also evaluate the impact of other LLMs in our experiments.
All experiments use surrogate data with 5000 sequences, i.e.5000 users.

\section{Validation Technique}
We propose a modified validation approach that better aligns with our objective of extracting models that closely mimic the target model's behavior. 
The original method in~\cite{yue2021black} considers only the validation item \(x_{[-2]}\) as ground truth, whereas the actual ground truth should be the entire recommendation list returned by the target model at that point.
Our validation metric computes NDCG by evaluating the surrogate model's top-\(k\) ranking \(L^k_s(\cdot)\) for DCG, where items in the target model's top-\(k\) recommendations \(\hat{L}^k(\cdot)\) have a relevance of 1, and others have 0. In this context, IDCG is calculated using the \(\hat{L}^k(\cdot)\):
\begin{equation}
    NDCG_k(L^k_s(\cdot), \hat{L}^k(\cdot))=\frac{DCG_k(L^k_s(\cdot))}{DCG_k(\hat{L}^k(\cdot))}.
\end{equation}

This differs from the original metric in~\cite{yue2021black}:
\[
\text{NDCG}(f_s(\cdot), \text{Sampler}(\hat{L}^k(\cdot))),
\]
where \(\text{Sampler}(f_t(\cdot))\) indicates a single item sampled from the target model's recommendations.

Our approach directly compares the top-\(k\) rankings between the surrogate and target models, providing a more accurate assessment of ranking agreement.

\section{Outlier Analysis}
Our method outperforms baselines in extracting NARM and SASRec across both data-free and data-limited settings. For BERT4Rec, while achieving the best results on Steam, performance on Beauty is sub-optimal, likely due to high sparsity (99.98\%) and weak sequential patterns. 
To validate this, we shuffle input sequences and compare top-100 recommendations. As shown in~\Cref{tab:sim}, BERT4Rec performs well without sequential cues on sparse datasets like Beauty and Games, explaining the strong MEA results with randomly generated sequences.
\begin{table}[]
\centering
\caption{Top-100 overlap ratio before and after input shuffling. Higher values imply weaker dependence on sequence order.}
\label{tab:sim}
\begin{tabular}{lccc}
\toprule
\tiny
& Beauty   & Games & Steam  \\  \midrule
NARM   & 74.01\% & 68.30\% & 70.39\%\\  
SASRec & 80.06\% & 66.51\% & 73.98\% \\  
BERT4Rec & 98.56\% & 98.24\% & 65.70\%  \\   
\bottomrule 
\end{tabular}
\end{table}

\section{Details of Model Architectures}
To evaluate the performance of attacks, we implement the model extraction attack on three representative sequential recommender models: NARM~\cite{li2017neural}, SASRec~\cite{kang2018self}, and BERT4Rec~\cite{sun2019bert4rec}. These models differ in their architectural components and training schemes, as summarized below:

\textbf{NARM}: This model integrates both global and local contextual information. It consists of an embedding layer, a gated recurrent unit (GRU) that functions as both a global and local encoder, an attention module to compute session features, and a similarity layer to output the most similar items to the session features. NARM is designed to capture long-term dependencies while effectively incorporating recent interactions.
    
\textbf{SASRec}: SASRec uses a stack of unidirectional transformer (TRM) layers to model sequential data. It begins with an embedding layer, incorporating both item and positional embeddings of the input sequence. Each transformer layer in SASRec contains a multi-head self-attention mechanism and a position-wise feed-forward network, which together enable SASRec to effectively capture sequential relationships in recommendation data.
    
\textbf{BERT4Rec}: Structurally similar to SASRec, BERT4Rec differs by using a bidirectional transformer and employing a masked language modeling (auto-encoding) task for training. This allows BERT4Rec to capture bidirectional dependencies in the input sequence, improving its performance by considering both past and future context in its recommendations.

\end{document}